 \documentclass[prb,aps,twocolumn]{revtex4}
\usepackage{graphicx} 
\usepackage{tabularx}
\usepackage{dcolumn} 
\usepackage{color}
\usepackage{bm}
\usepackage{amsmath}
\usepackage{amssymb}

\begin{document} 


\title{Theory of spin-polarized superconductors
--an analogue of superfluid $^3$He A-phase--
}

\author{Kazushige Machida} 
\affiliation{Department of Physics, Ritsumeikan University, 
Kusatsu 525-8577, Japan} 

\date{\today}

\begin{abstract}
It is shown theoretically that ferromagnetic superconductors,
 UGe$_2$, URhGe, and UCoGe can be described in terms of the A-phase like triplet pairing 
 similar to superfluid $^3$He in a unified way, including peculiar reentrant, S-shape, or L-shape $H_{\rm c2}$ curves.
 The associated double transition inevitable between the A$_1$ and A$_2$-phases in the $H$-$T$
 plane is predicted, both of which are characterized by non-unitary state with broken time reversal symmetry
 and the half-gap. UTe$_2$, which has been discovered quite recently to be a spin-polarized superconductor,
 is analyzed successively in the same view point, pointing out  that the expected 
 A$_1$-A$_2$ transition is indeed emerging experimentally. Thus the four heavy Fermion compounds all together
 are entitled to be topologically rich solid state materials worth further investigating together with superfluid $^3$He A-phase.
 \end{abstract}

 
 
\maketitle 

Much  attention has been focused on ferromagnetic superconductors~\cite{aoki}, such as UGe$_2$, URhGe, and UCoGe.
Recently a new member of a superconductor UTe$_2$ with $T_{\rm c}$=1.6K~\cite{aoki2,ran} is discovered, 
which is almost ferromagnetic, and attracts much excitement. 
Many researchers start devoting to UTe$_2$ experimentally~\cite{knebel,daniel,miyake,ran2,metz,mad,tokunaga,sonier,nakamine,1,2,3,4} and 
theoretically~\cite{xu,ishizuka,shick} and
show renewed interest on the former three compounds too.
Those heavy Fermion materials belong to a strongly correlated system
where 5f electrons responsible for it is believed to form a coherent narrow band with the large mass enhancement below a  Kondo temperature.
Since $H_{\rm c2}$ in those compounds exceeds the Pauli paramagnetic limitation,
it is thought that a triplet or odd parity pairing state is realized~\cite{aoki}.
However, the detailed studies of the pairing symmetry remains lacking in spite of the long history of the first three compounds
over two decades.
Some of those features are, interestingly enough, common for spin-polarized flat band superconductivity found in 
double bilayer magic angle graphene~\cite{liu,ashvin},
where the in-plane $H_{\rm c2}$ shows a similar ``reentrance'' behavior, as will see shortly,
due to linear Zeeman effect in the superconducting (SC) state, appearing next to a spin-polarized insulating state.

To understand those four spin-polarized superconductors in a unified way, 
here we  develop a phenomenological theory based on
the assumption that the four compounds are commonly described 
in terms of the triplet pairing symmetry analogous to the superfluid 
$^3$He-A phase~\cite{leggett} under Zeeman effect~\cite{mermin}.
Namely the A$_1$-A$_2$ phase transition is induced by an applied field, 
which is observed as the clear double specific heat jumps~\cite{halperin}.

There exist prominent SC properties observed commonly in those superconductors:

\noindent
(1) Above  $T_{\rm c}$ the ferromagnetic transition (FM) occurs in the first three compounds.
Thus the SC state survives under a strong internal field coming through an exchange interaction. 
However, in UTe$_2$ ``static'' FM is not detected so far 
although FM fluctuations are probed~\cite{ran,miyake,tokunaga,sonier} above  $T_{\rm c}$, i.e.
a diverging static susceptibility along the $a$-axis~\cite{ran,miyake}, and $1/T_2$~\cite{tokunaga}. 

\noindent
(2) When $H$ is applied parallel to the magnetic hard axis $b$ in orthorhombic  crystals, $H_{\rm c2}$ exhibits
the reentrant behavior for URhGe where the SC state once disappeared reappears
again at higher fields, or an S-shape  $H_{\rm c2}$ in UCoGe~\cite{aoki} and a L-shape in UTe$_2$~\cite{knebel} 
in the $H$-$T$ plane.

\noindent
(3) The gap structure is unconventional characterized by either point  in UTe$_2$~\cite{aoki2,ran} or line 
nodes in the other three~\cite{aoki}.

\noindent
(4) There is clear experimental evidence for double transitions:
the two successive second order phase transitions seen by  specific heat experiments as distinctive jumps systematically change 
under pressure ($P$) in UTe$_2$~\cite{daniel}.
A similar indication for the double transitions in the ambient pressure is found in UCoGe at $T_{\rm c2}\sim$0.2K~\cite{manago}
where $1/T_1T$ exhibits a plateau corresponding to the ``half residual DOS'' value at the intermediate $T$ below $T_{\rm c}=$0.5K.
Upon further lowering $T$ it stars decreasing again at 0.2K.

\noindent
(5) In UTe$_2$ under ambient pressure a quite similar $1/T_1T$ is also observed in NMR experiment~\cite{nakamine}:
The saturated $1/T_1T=0.25$ normalized by the normal value, which corresponds to the ``half residual DOS'', restarts decreasing to zero
at the second transition at $T_{\rm c2}$=0.2K. This value nicely matches with the missing second
phase transition in the $P$-$T$ plane, namely, a smooth extrapolation of $T_{\rm c1}$ and $T_{\rm c2}$ toward $P\rightarrow 0$.

\noindent
(6) Recent specific heat $C/T$ data for several high quality samples of UTe$_2$~\cite{aoki2,ran,shimizu} 
commonly show the residual DOS amounting to $0.5N(0)$, a half of the 
normal DOS $N(0)$ while some claims zero~\cite{metz}. 
Thus this ``residual'' half DOS issue is controversial at this moment.
We will propose to resolve it later in this paper.

In order to address those seemingly ``complicated and mutually conflicting'', but quite intriguing experimental facts, 
we postulate the A phase like triplet pair symmetry which responds to the spontaneous 
FM moment under perpendicular external fields to yield the A$_1$-A$_2$ double transitions.
This scenario coherently explains the observed reentrant $H_{\rm c2}$ in URhGe, an S-shape in UCoGe, and L-shape in UTe$_2$
in a unified way.

As mentioned, the A$_1$-A$_2$ phase transition in $^3$He A-phase~\cite{halperin} is controlled by the linear 
Zeeman effect due to applied field, which acts to split $T_{\rm c}$~\cite{mermin}. 
Thus in the FM superconductors $T_{\rm c}$ is controlled by 
spontaneous magnetic moment of ferromagnetism, which is linearly coupled to the non-unitary order parameter. 
We apply a Ginzburg-Landau theory to describe those characteristic $H_{\rm c2}$ curves.
 We also identify the pairing symmetry based on group theoretic classification~\cite{machida}. The 
pairing symmetry is non-unitary triplet~\cite{machida,machida2,ohmi}, 
where the {\bf d}-vector is a complex function and points to the perpendicular direction to the magnetic 
easy axis and the gap function has either point or line nodes with possibly chiral $p$-wave orbital form, 
which is most consistent with the STM observation of chiral edge states~\cite{mad}

Let us start with  Ginzburg-Landau theory for the A-phase like pairing state described by the complex 
${\bf d}$-vector ${\bf d}(k)=\phi(k){\vec \eta}$ with ${\vec \eta}={\vec \eta}'+i{\vec \eta}''$ 
(${\vec \eta}'$ and ${\vec \eta}''$ are real vectors).
$\phi(k)$ is the orbital part of the pairing function which is classified group-theoretically under SO(3)$\times$D$_{2h}$ 
symmetry~\cite{machida,annett}.
Here we assume the weak spin-orbit coupling scheme whose strength depends on the compounds
and will be appropriately tuned relative to the experimental situations.
There exists U(1)$\times$Z$_2$ symmetry in this pairing, namely invariance
under ${\bf d}\rightarrow -{\bf d}$ and the gauge transformations.

Under D$_{2h}$ symmetry GL free energy functional is given
by

\begin{eqnarray}
\label{e1}
F=\alpha_0(T-T_{\rm c0}){\vec \eta}\cdot{\vec \eta}^{\star}+{\beta_1\over2}({\vec \eta}\cdot{\vec \eta}^{\star})^2
+{\beta_2\over2}|{\vec \eta}^2|^2 \nonumber \\
+i\kappa {\vec M}\cdot {\vec \eta}\times {\vec \eta}^{\star}.
\end{eqnarray}

\noindent
The last invariant ($\kappa>0$) comes from the non-unitarity of the pairing function in the presence of the 
spontaneous moment $\vec M(H)$, which responds to external field directions differently.
Since the fourth order term are written as 
$F^{(4)}={\beta_1\over2}({\vec \eta}'\cdot{\vec \eta}'+{\vec \eta}''\cdot{\vec \eta}'')^2
+{\beta_1\over2}[({\vec \eta}'\cdot{\vec \eta}'-{\vec \eta}''\cdot{\vec \eta}'')^2+4({\vec \eta}'\cdot{\vec \eta}'')^2]$,
for $\beta_1, \beta_2>0$, we can find the minimum when $|{\vec \eta}'|=|{\vec \eta}''|$
and ${\vec \eta}'\perp{\vec \eta}''$.
Note that the weak coupling estimate leads to 
${\beta_1\over\beta_2}=-2$, thus we assume the strong coupling effects in the following arguments.

The quasi-particle spectra are given by
$E_{k,\sigma}=\sqrt{\epsilon(k)^2+(|{\vec \eta}|^2\pm|{\vec \eta}\times{\vec \eta}^{\star}|)\phi(k)^2}$.
If we choose ${\vec \eta}'=\eta_b {\hat b}$ and ${\vec \eta}''=\eta_c {\hat c}$, those are rewritten as
$E_{k,\sigma}=\sqrt{\epsilon(k)^2+\Delta_{\sigma}(k)^2}$
where the gap functions for two branches are $\Delta_{\uparrow}(k)=|\eta_b+\eta_c|\phi(k)$
and $\Delta_{\downarrow}(k)=|\eta_b-\eta_c|\phi(k)$.
Note that 
(1) If $|\eta_c|=0$, $\Delta_{\uparrow}(k)=\Delta_{\downarrow}(k)$, which is the A phase~\cite{leggett}.
(2) When $|\eta_b|=|\eta_c|$, $\Delta_{\uparrow}(k)\neq0$ and $\Delta_{\downarrow}(k)=0$
which is the non-unitary A$_1$ phase~\cite{mermin}. The gap of one of the two branches identically vanishes and remains normal.
Therefore, if we naively assume that in the normal state $N_{\uparrow}(0)=N_{\downarrow}(0)$
which is consistent with the NMR data in UTe$_2$~\cite{nakamine},
the A$_1$ phase is characterized by having the ungapped DOS $N_{\downarrow}(0)=N(0)/2$
with $N(0)=N_{\uparrow}(0)+N_{\downarrow}(0)$.
In the non-unitary state with the complex $\bf d$-vector,
the time reversal symmetry is broken.

It is convenient to consider ${\vec \eta}=(0,\eta_b,\eta_c)$ or $\eta_{\pm}={1\over \sqrt2}(\eta_b\pm i\eta_c)$ 
for ${\bf M}=(M_a,0,0)$. Then we see from eq. (\ref{e1}), the quadratic term $F^{(2)}$ becomes

\begin{eqnarray}
F^{(2)}=\alpha_0(T-T_{\rm c \uparrow})|\eta_{+}|^2+\alpha_0(T-T_{\rm c \downarrow})|\eta_{-}|^2
\label{e2}
\end{eqnarray}

\noindent
with $T_{\rm c \uparrow, \downarrow}=T_{\rm c0} \pm{\kappa\over \alpha_0}M_a$.
The actual second transition temperature is modified to 
$T_{\rm c \downarrow}=T_{\rm c0}-{\kappa M_a \over \alpha_0} \cdot{{\beta_1-\beta_2}\over{2\beta_2}}$,
which could be larger or smaller than the original $T_{\rm c \downarrow}^{(0)}\equiv T_{\rm c0}-(\kappa/\alpha_0)M_a$
due to the fourth order terms. For $1\le{\beta_1/\beta_2}\le3$, 
$T_{\rm c \downarrow}>T_{\rm c \downarrow}^{(0)}$.
This remark becomes important to understand the asymmetric L-shape $H_{\rm c 2}^b$ observed in UTe$_2$~\cite{knebel}
as see later.
It is easy to see the ratio of the specific heat jumps
${\Delta C(T_{\rm c \uparrow})\over \Delta C(T_{\rm c \downarrow})}={T_{\rm c \uparrow}\over 
T_{\rm c \downarrow}}\cdot {\beta_1\over{\beta_1+\beta_2}}$.
The jump at $T_{\rm c \downarrow}$ can be quite small for $T_{\rm c \uparrow}\gg T_{\rm c \downarrow}$ which is the case for 
UTe$_2$ since $T_{\rm c \uparrow}$=1.6K, $T_{\rm c \downarrow}$=0.2K, and the second factor is an order of one,
anticipating the difficulty to observe the second transition thermodynamically.

The FM moment $M_a$ acts to shift the original transition temperature $T_{\rm c0}$ and split it into $T_{\rm c \uparrow}$
and $T_{\rm c \downarrow}$. The external field $H$ comes in through $M_a(H)$ in addition to usual vector potential
which gives rise to orbital depairing.
The magnetic coupling is estimated~\cite{mermin} by
$\kappa=T_{\rm c}{N'(0)\over N(0)}ln(1.14\omega/T_{\rm c})$
where $N'(0)$ is energy derivative of the normal DOS and $\omega$ energy cutoff.
This term arises from the electron-hole asymmetry near the Fermi level. $\kappa$ indicates
the degree of this asymmetry, which can be substantial for a narrow band, thus the Kondo coherent band in heavy Fermion
material of our cases is expected to be important.
We can estimate
$N'(0)/N(0)\sim 1/E_{\rm F}$ with the Fermi energy $E_{\rm F}$.
Since $T_{\rm c}$=2mK and $E_{\rm F}$=1K in $^3$He, $\kappa\sim10^{-3}$,
while the present compounds $T_{\rm c}\sim$1K and $E_{\rm F}\sim T_{\rm K}$ with the $T_{\rm K}$ 
Kondo temperature typically 10$\sim$50K,
thus  $\kappa$ is an order of $10^{-1}$.

Let us now consider the action of external field $H_b$ applied to the hard axis $b$ on the FM moment $M_a$, 
pointing parallel to the $a$-axis.
The $a$-axis component of the moment $M_a(H_b)$ generally decreases by rotating it toward the $b$-axis as shown in Fig. 1(b).
In fact it is actually observed in URhGe by neutron experiment~\cite{huxley}. Here we display the generic and typical magnetization curves of $M_a$
and $M_b$ in Fig. 1(c) where $H_R$ denotes a characteristic field for $M_b(H_b)=M_a$.
Experimentally it is realized by a meta-magnetic transition via a first order transition in UTe$_2$~\cite{miyake} or 
gradual change in UCoGe~\cite{aoki}.

Thus as displayed in Fig.1(a), $T_{\rm c \uparrow}$ ($T_{\rm c \downarrow}$) by increasing 
$H_b$ decreases (increases) according to eq. (\ref{e2}). The two transitions $T_{\rm c \uparrow}$ =$T_{\rm c \downarrow}$
meet at $H_R$. Upon further increasing $H_b$, $T_{\rm c \downarrow}$ could keep increasing further
by rotating the $\bf d$-vector direction so that $\bf d$ is now perpendicular to ${\bf M}_b$, 
which maximally gains the magnetic coupling energy
$i\kappa {\vec M}\cdot {\vec \eta}\times {\vec \eta}^{\star}$.
This process occurs gradually or suddenly, depending on the situations of the magnetic subsystem
and also on the spin-orbit coupling which locks $\bf d$ to underlying lattices.
Therefore $H_R$ may indicate simultaneously the $\bf d$-vector rotation.
It should be noted, however, that if the spin-orbit coupling is so strong, the $\bf d$-vector rotation is prevented.
In this case $H^b_{\rm c2}$ exhibits a Pauli limited behavior as observed in UTe$_2$ under pressure~\cite{aoki3}.

\begin{figure}
\includegraphics[width=8.5cm]{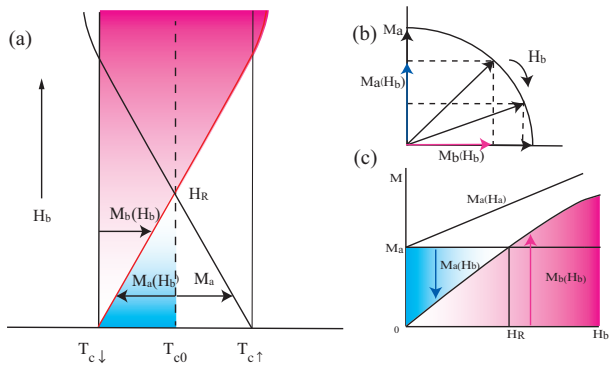}
\caption{(a) Generic phase diagram of the two transition temperatures $T_{\rm c \uparrow}$ and $T_{\rm c \downarrow}$
under applied field $H_b$. At $H_b$=0,  $T_{\rm c \uparrow}$ and $T_{\rm c \downarrow}$ are split from 
$T_{\rm c0}$ in proportion to $M_a$. As $H_b$ increases, $T_{\rm c \uparrow}$ ($T_{\rm c \downarrow}$)
decreases (increases) because $M_a(H_b)$ diminishes and meet at $H_R$. By rotating the $\bf d$-vector 
$T_{\rm c \downarrow}$ keeps increasing.
(b) By increasing $H_b$ the spontaneous moment $M_a$ rotates toward the $b$-axis.
The projection of $M_a(H_b)$ to the $a$-axis decreases, and it vanishes at $H_R$ where the moment completely points to 
the $b$ direction. (c) Schematic magnetization curves for the spontaneous moment $M_a$ and the induced moment $M_b$.
At $H_R$, $M_b(H_b)$ becomes equal to $M_a$ for $H$=0.
}
\label{f1}
\end{figure}

Within the GL scheme it is easy to estimate $H^b_{\rm c2}$ as follows.
We start with the $H^b_{\rm c2}$ expression:
$H_{\rm c2}(T)=A_0(T_c(H_{\rm c2})-T)$
with $A_0={\Phi_0\over 2\pi\hbar^2}4m\alpha_0$, $m$ effective mass, and $\Phi_0$ quantum unit flux.
 Here $T_c$ depends on $H$ though $M_a(H)$  as described above.
Thus the initial slope of $H'_{\rm c2}$ at $T_c$ is simply given by
$H'_{\rm c2}(T)=A_0{dT_c\over dH_{\rm c2}}H'_{\rm c2}-A_0$.
It is seen that if ${dT_c\over dH_{\rm c2}}=0$ for the ordinary superconductors, $H'^0_{\rm c2}(T)=-A_0<0$. The slope 
$H'_{\rm c2}(T)$ is always negative.
However, it is easily written as


\noindent

\begin{equation}
{1\over |H'_{\rm c2}|}={1\over |H'^0_{\rm c2}|}+|{dT_c \over dH_{\rm c2}}|.
\end{equation}

\noindent
The condition for attaining the positive slope, $H'_{\rm c2}(T)>0$ implies $|H'^0_{\rm c2}|>({{dH_{\rm c2}}\over dT_c})$.
This is a necessary condition to achieve an S-shaped or L-shaped $H_{\rm c2}$ curves.
This is fulfilled when $|H'^0_{\rm c2}|$ is large enough, that is, the orbital depairing is small 
or $|{dT_c\over{dH}}|$ at $H_{\rm c2}$ is large, or the $T_c$ rise is strong enough.
It is to be noted that when $1-A_0({dT_c\over dH_{\rm c2}})=0$, the $H_{\rm c2}(T)$ curve has a 
divergent part, which is indeed observed in UCoGe as a part of the S-shape. 
It is clear from the above that when $dT_c/dH<0$,
$|H'_{\rm c2}(T)|<|H'^0_{\rm c2}|$.
Namely, in this case the slope $|H'_{\rm c2}|$ is always smaller than the original $|H'^0_{\rm c2}|$
as expected.

\begin{figure}
\includegraphics[width=9cm]{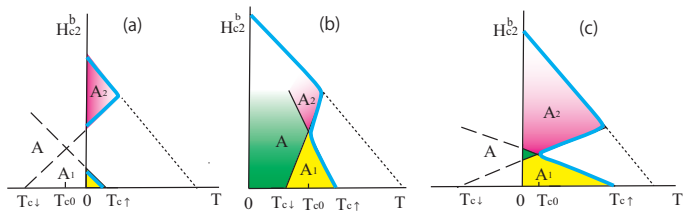}
\caption{Schematic possible three phase diagrams in $H( \parallel b)$ and $T$ plane.  $H_{\rm c2}^b$ exhibits
the separated reentrance as in URhGe (a),  S-like curve as in UCoGe (b), and L-like shape as in UTe$_2$ (c).
The dotted lines show the absolute upper critical field $H_{\rm c2}^{\rm upper}$.  
Beyond this domain there exists no SC state allowed.
}
\label{f1}
\end{figure}

When the magnetic field is applied to the magnetic easy $a$-axis, the spontaneous moment $M_a(H_a)$
increases monotonically as shown in Fig.1(c).  According to eq. (2), $T_{c\uparrow}$ ($T_{c\downarrow}$)
increases (decreases) as $H_a$ increases. Thus $H_{c2}^{\uparrow}$ can have a positive slope at  $T_{c\uparrow}$
in principle. However, the existing data in UCoGe~\cite{wu} show that it is negative and there is no report for it in other compounds.
This is simply because the
orbital depairing $H'^0_{\rm c2}$ overcomes the positive rise of $T_{c\uparrow}$.
 Moreover, $H_{c2}^{\downarrow}$ is strongly suppressed by both $T_{c\downarrow}$ and the orbital
 effect $H'^0_{\rm c2}$, resulting in quite a low $H^a_{c2}$, compared with $H^b_{c2}$.
 This large $H_{c2}$ anisotropy is common in those compounds~\cite{aoki}.
 From the above calculation, it is apparent that the large $H^b_{c2}$ comes because the
 higher field part of $H_{c2}$ belongs to $H_{c2}^{\downarrow}$, which has a positive slope.

There could be several types of the $H^b_{\rm c2}$ curves depending on the magnitude of the spontaneous 
moment $M_a$, and its growth rate against $H_b$, the coupling constant $\kappa$, $T_{\rm c0}$, and etc.  
Possible representative $H^b_{\rm c2}$ curves are displayed in Fig. 2.

When the hypothetical $T_{\rm c0}$ is situated in the negative temperature side, the realized phase is the A$_1$-phase only at $H_b=0$.
In the high field regions, $T_{\rm c \downarrow}(H_b)$ appears by increasing $M_b(H_b)$ as the A$_2$-phase, which is shown in Fig. 2(a).
This is the situation of URhGe where the reentrant phase appears around the magnetic rotation field $H_R$~\cite{aoki}.
We also note that under uniaxial stresses along the $b$-axis $H_{\rm c2}^b$ curve continuously deforms
and two separated SC regions merge into an S-shaped $H_{\rm c2}^b$ in URhGe (see Fig. 2(a) in Ref.~\onlinecite{daniel2}).
This can be understood because under stresses $H_R$ is known to decrease, thus 
the detached high field part approaches and eventually touches the lower part to from an S shape $H_{\rm c2}^b$,
resulting in a positive slope $(dH_{\rm c2}^{b}/dT)_{T_{\rm c}} >$0.

The second example shown in Fig. 2(b) is the case where $M_a$ and $\kappa$ are relatively small,
thus the splitting $T_{\rm c \uparrow}$ and $T_{\rm c \downarrow}$ at $H_b=0$ is small. The small moment $M_a=0.05 \mu_{\rm B}$
and a rapid growth of $M_b$ in UCoGe~\cite{aoki} are consistent with this picture.
The resulting $H^b_{\rm c2}$ curve
possess an S shape, similar to that observed in UCoGe.
Note that there must always exist the ``absolute'' upper limit of $H_{\rm c2}$ for any superconductors. This is set by the orbital depairing
where $H_{\rm c2}^{\rm upper}= {\Phi_0/2\pi\xi^2}$ at $T$=0  with $\xi$ coherent length.
This absolute upper limit $H_{\rm c2}^{\rm upper}$ has a triangle region in the $H$ and $T$  plane which is denoted by a dotted line in Fig. 2.
The SC state is allowed only inside this $H_{\rm c2}^{\rm upper}$ curve.
If the rising $T_{\rm c \downarrow}(H_b)$ hits this line, the $H_{\rm c2}$ curve follows this limiting curve.

In Fig. 2 (c), we show the $H^b_{\rm c2}$ curve similar to UTe$_2$ with an L shape~\cite{knebel}.
It occurs when $H_R$ situates at lower field and $T_{\rm c0}$ is positive and low.
All three phases A$_1$, A$_2$ and A are realized for $H\parallel b$-axis.
The observed $H_{\rm c2}^b$ curve in UTe$_2$ is different from Fig.2 (c) in that
the upper part of the A$_2$-phase is horizontally cut out~\cite{knebel} at $H_{\rm c2}^b\sim$40 T because beyond that field the first order 
meta-magnetic transition completely alters the background electronic state, so that the
SC state is abruptly wiped out. However, note that the field direction is
appropriately chosen away from the $b$-axis, $H_{\rm c2}$ is greatly enhanced by avoiding the meta-magnetic transition
at $H\parallel b$, reaching $\sim$60T~\cite{ran2}.
This means that the intrinsic absolute upper limit of the orbital  $H^{upper}_{\rm c2}$ 
can reach remakably $\sim60$T in this system.
Thus it was quite reasonable to ignore the orbital depairing effect  in the above $H_{\rm c2}$ arguments.

Therefore, it is possible to reproduce the essential features
associated with $H_{\rm c2}$ for all compounds in terms of the A$_1$-A$_2$
transitions. To confirm our scenario, we now examine possible signatures
 of the double transition. The recent NMR experiment~\cite{nakamine} on UT$_2$ clearly demonstrates it 
 at $T_{\rm c\downarrow}\sim0.2$K. Namely, upon lowering $T$, $1/T_1T$ which is proportional to square of DOS
 exhibits a sudden drop at $\sim0.2$K after a prolonged $T$-constant plateau at 0.25 normalized by its normal value
 corresponding to $N(0)/2$. This behavior is backed up by the simultaneous Knight shift measurement~\cite{nakamine}.
 $T_{\rm c\downarrow}=0.2$K moves up from $T_{\rm c\downarrow}^{(0)}$ which is expected to situate at a negative $T$
 as shown in Fig. 2(c) because of the fourth order terms in eq. (1) mentioned before.
 
Under $P$ in UT$_2$ the successive double transitions are indeed discovered~\cite{daniel}
and vary systematically as explained in the $P$-$T$ phase diagram. The missing second transition at 
ambient $P$ is now found mentioned above.

As for UCoGe, thermal conductivity experiment~\cite{taupin} indicates an anomaly at $T\sim0.2K$,
which coincides roughly with our prediction shown in Fig. 2(b). As a function of $H\parallel b$
the thermal conductivity anomaly is detected as a sudden increase at $H\sim0.6H_{\rm c2}$ 
(see Fig. 5 in Ref.~\onlinecite{wu1}). Moreover, under $H$ parallel to
the easy axis $H_{\rm c2}$ curve shows low $T$ enhancement indicative of the underlaying phase 
transition (see Fig. 2(b) in Ref.~\onlinecite{wu}). 
According to NMR by Manago et al~\cite{manago},  $1/TT_1$ shows very similar behaviors: plateau at $N(0)/2$ and
sudden drop mentioned above. We expect further careful experiments to detect the
A$_1$-A$_2$ transitions in all four compounds.

While for UGe$_2$, URhGe, and UCoGe ``static" FM transitions are established, in UTe$_2$
slow FM fluctuations are found~\cite{ran,miyake,tokunaga,sonier}, which could be the origin of the symmetry breaking of
$T_{\rm c\uparrow}\neq T_{\rm c\downarrow}$ under the assumption that FM fluctuations 
are slow enough compared with the conduction electron motion.
We recall a similar circumstance in UPt$_3$:  The fluctuating antiferromagnetism
at $T_N=5$K is detected only by the fast probe: neutron diffraction~\cite{aeppli,trappmann} and undetected by
other ``static'' probes, such as specific heat, $\mu$SR, and NMR.
Yet, this is believed to be the origin of the double transition in UPt$_3$~\cite{UPt3,sauls}.

In summary, we have discussed the SC properties of UGe$_2$, URhGe, UCoGe, and UTe$_2$ in depth
in terms of a non-unitary triplet pairing state in a unified way. The FM moment
governs and produces the various types of the $H_{\rm c2}$ curves observed.
The possible pairing function is described by the complex ${\bf d}$-vector whose direction is perpendicular to
the magnetic easy axis. The orbital part $\phi(k)$ of the order parameter could be line nodes allowed group-theoretically~\cite{machida} in 
the present orthorhombic symmetry, which is experimentally suggested in UCoGe~\cite{aoki,taupin,wu}.

As for UTe$_2$, the specific heat experiments~\cite{ran,aoki2,metz} exhibit  $C/T\sim T^2$, 
suggesting the gap structure with
point nodes, which is also consistent with microwave measurement~\cite{1}. 
Thus if this is true, the pairing function of UTe$_2$ is symbolically given by $({\bf b}+i{\bf c})(k_b+ik_c)$,
a solid state analogous literally to superfluid $^3$He A$_1$-phase~\cite{mizushima},  which is doubly
chiral both in the spin and orbital parts with points nodes. 
This chiral $p$-wave form in a simplest, but somewhat ``ad hoc'' possibility
is consistent with the observed chirality by STM~\cite{mad}.
The orbital angular moment ${\bf L}$ spontaneously induced by this
chiral state can gain the extra energy through the coupling ${\bf M}\cdot{\bf L}$ with the spontaneous magnetic moment.
We can explore a variety of interesting topological properties, such as Weyl nodes associated with point nodes, 
known in $^3$He A-phase~\cite{mizushima} which was difficult to access experimentally.

The author is grateful for useful discussions with Y. Shimizu, Y. Tokunaga, A. Miyake, T. Sakakibara, S. Nakamura, S. Kittaka,
M. Manago, S. Kitagawa, and G. Nakamine. He especially 
thanks D. Aoki and K. Ishida for sharing the data prior to publication. This work is supported by JSPS KAKENHI, No. 17K05553
and partly performed at the Aspen Center for Physics, which is supported by National Science Foundation grant PHY-1607611.

\end{document}